\documentclass[aps,prl,superscriptaddress,showpacs,twocolumn]{revtex4}
\usepackage{epsfig,graphicx,psfrag,amsmath}
\begin{document}
\title{Full counting statistics of strongly non-Ohmic transport through single molecules}
\author{Jens Koch}
\affiliation{Institut f\"ur Theoretische Physik, Freie Universit\"at Berlin, Arnimallee
14, 14195 Berlin, Germany}
\author{M.E.\ Raikh}
\affiliation{Department of Physics, University of Utah, Salt Lake City, 
UT 84112}
\author{Felix von Oppen}
\affiliation{Institut f\"ur Theoretische Physik, Freie Universit\"at Berlin, Arnimallee
14, 14195 Berlin, Germany}
\date{August 4, 2005}
\begin{abstract}
We study analytically the full counting statistics of charge transport
through single molecules, strongly coupled to a weakly damped
vibrational mode. The specifics of transport in this regime -- a
hierarchical sequence of avalanches of transferred charges,
interrupted by ``quiet" periods -- make the counting statistics
strongly non-Gaussian. We support our findings for the counting
statistics as well as for the frequency-dependent noise power by
numerical simulations, finding excellent agreement.

\end{abstract}

\maketitle

{\it Introduction.}---A prime qualitative difference of transport
through single molecules as compared to artificial nanostructures lies
in the role of the vibrational motion of the nuclei. This aspect is at
the focus of current experiments \cite{Parks,Ruitenbeek,
Ralph}, and is also being studied theoretically
within a number of approaches \cite{Glazman,Schoeller,Flensberg,Aleiner,Nitzan,Takei,Koch}. The
incorporation of molecular vibrations (phonons) into the
theoretical description is mostly done within simplified
(phenomenological) models, as opposed to purely electronic
first-principles studies \cite{Ratner,Baranger}.

Even within minimal models involving one molecular orbital coupled to
a single vibrational mode, ``unidirectional'' transport (i.e., for voltages 
large compared to temperature) depends radically on the strength
of the electron-phonon coupling, already at the qualitative level \cite{Flensberg,Aleiner,Koch}. 
For weak and intermediate coupling \cite{terminology}, transport is adequately described in terms of
individual electron transitions. By contrast when vibrational relaxation is slow, 
transport in the regime of strong electron-phonon coupling is appropriately captured within 
a scenario of {\it avalanches} of transferred electrons,
with exponential spreads of height and duration \cite{Koch}. 

More quantitatively, the time dependence of the current in the strong-coupling regime
can be presented as 
\begin{equation}
\label{current}
  I(T) = f_1^{(0)}(T-t_1) + f_2^{(0)}(T-t_1-t_2) + \ldots, 
\end{equation}
where $t_i$ are the time intervals between avalanches (quiet periods). 
These intervals are much longer than the typical duration $\tau^{(0)}$
of an avalanche which occurs during the sparse periods when the vibrations 
are excited. The random function $f_i^{(0)}(\tau)$ (which is nonzero only for
times $|\tau|\lesssim \tau^{(0)}$) describes the passage of a {\em large number} 
$\int d\tau f_i^{(0)}(\tau)=N_i \gg 1$ of electrons through the molecule during the $i$th avalanche.
Moreover, a numerical study of the avalanches \cite{Koch} revealed their self-similar 
hierarchical structure, see Fig.\ 1. Quantitatively, this structure manifests
itself in the fact that, during the time of an avalanche $\sim \tau^{(0)}$, 
each function $f_i^{(0)}$ itself takes the form of Eq.\ (\ref{current}),
with $f_i^{(0)}$ replaced by random functions $f_i^{(1)}$, which describe
avalanches of the {\em first generation} interrupted by quiet periods.
Again, these quiet periods are much longer than the characteristic time scale $\tau^{(1)}$
of the functions $f_i^{(1)}$. For times shorter than  $\tau^{(1)}$, the  
functions $f_i^{(1)}$ have the form of Eq.\ (\ref{current}) with corresponding 
second-generation avalanches, $f_i^{(2)}$, having even shorter time-scale, $\tau^{(2)}$,
and so on \cite{cutoff}. Numerical results supporting this scenario are shown in Fig. 1.

The above discussion implies that the statistical
properties of charge transport through a molecule in the regime of
strong electron-phonon coupling and through a conventional nanostructure
are drastically different. For a nanostructure, all $f_i^{(0)}$
are $\delta$ functions, so that $N_i=1$. Hence, the distribution function 
$P_T(Q)$ of the net transmitted charge $Q$ during time $T$  (full counting 
statistics \cite{Lesovik}) is completely encoded in the distribution of the 
{\em waiting times} $t_i$ for single-electron transitions. 
This distribution reflects
the details of the transport mechanism, and might be quite nontrivial \cite{Blanter}. 
Nevertheless, with all $t_i$ being of the same order, the full 
counting statistics differs only weakly from a Gaussian distribution. 
Small deviations are caused by correlations \cite{Levitov,Imry}, 
interactions \cite{Kindermann}, or the influence of the environment \cite{Beenakker},
and have been extensively studied theoretically.

By contrast, the counting statistics of avalanche-type transport is
{\em insensitive} to the details of the passage of a single
electron through the molecule, since the number of 
electrons  involved in each avalanche is large. 
Instead, the counting statistics is governed {\em exclusively} by 
the transition rates between different vibrational states. 
These rates have a simple structure in the limit of strong coupling
which allows us to develop a complete analytical theory for the regime of 
avalanche-type transport. In particular, we demonstrate in this paper 
that the full counting statistics $P_T(Q)$ is given by a  
concise analytical expression, which is strongly skewed
at ``short'' times ($\sim$ zero-order quiet period)    
and evolves into a Gaussian only for very large $T$.
Along with the counting statistics, we also study
analytically how the hierarchy of avalanches manifests itself in the
frequency dependence of the noise power $S(\omega)$. 
Our analytical results are in excellent agreement with numerical 
Monte-Carlo (MC) simulations.

\begin{figure}[t]
\includegraphics[viewport=50 620 310 800,width=\columnwidth,clip]{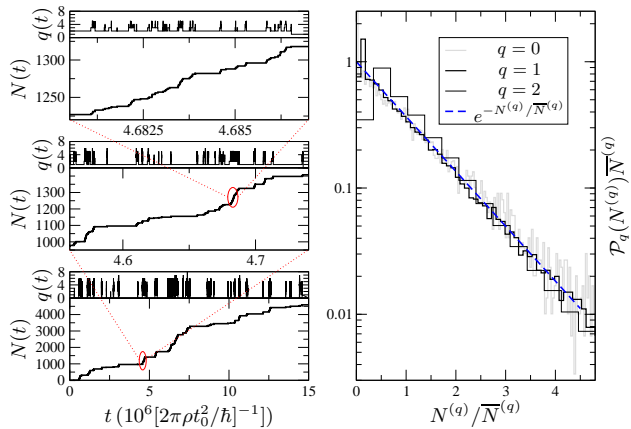}
\caption{Hierarchical character of transport. Left: Three generations of self-similar MC plots for $\lambda=4$ and $eV=3\hbar\omega_v$, showing the net-transferred charge $N$ and phonon state $q$ as functions of time. 
($\omega_v$: phonon frequency; $\rho$: density of states of the leads; $t_0$: molecule-lead
coupling).
Right: Comparison of the fixed-point distribution for the transferred charge per generation-$q$ avalanche 
to numerical simulations for $q=0,1,2$  (mean values $\overline{N}^{(0)}=91.2$, $\overline{N}^{(1)}=11.1$, and $\overline{N}^{(2)}=2.9$). 
}
\label{montecarlo}
\end{figure}

{\it Full counting statistics.}---Since different zeroth-generation avalanches are statistically independent,
it is easy to derive a relation between the counting statistics $P_T(Q)$ of the 
{\em net charge} $Q$ and the conventional counting statistics ${\varphi}_T(n)$ \cite{Lesovik}  
of the {\em number} of zeroth-generation avalanches $n$ during the time interval $T$. Indeed, using the definition  
$P_T(Q)=\langle \delta(Q - \sum_{j=1}^n N_j) \rangle_{N_j,n}$, and a 
Fourier representation of the RHS, one obtains
\begin{equation}
\label{general}
P_T(Q)=\int \frac{d\alpha}{2\pi}e^{i\alpha Q}\sum_n \left[\tilde{\cal P}_0(\alpha)\right]^n \varphi_T(n),
\end{equation}
where $\tilde{\cal P}_0(\alpha)=\langle\exp\left(-i\alpha N_j\right)\rangle_{N_j}$ denotes the Fourier
transform of the distribution function ${\cal P}_0(N)$ of the total charge passing per 
zeroth-generation avalanche. The durations of the quiet periods obey Poisson statistics so that
$\varphi_T(n) = \exp(-{\overline n}_T){\overline n}_T^n/n!$. Here, ${\overline n}_T$ denotes 
the average number of zeroth-generation avalanches within time $T$. 
Substituting this form into Eq.\ (\ref{general}) and performing the summation over $n$ yields 
an expression for the  counting statistics similar to the Holtsmark distribution \cite{Holtsmark},
\begin{equation}
\label{similar}
   P_T(Q) = \int \frac{d\alpha}{2\pi} \exp\left\{ i\alpha Q + {\overline n}_T \left[\tilde{\cal P}_0(\alpha) 
   - 1  \right]\right\}.
\end{equation}
Thus, the problem of the counting statistics is reduced to finding the distribution
${\cal P}_0(N)$. Two facts allow us to find ${\cal P}_0(N)$, namely ({\em i}) 
the self-similar structure of  avalanches and  ({\em ii}) the large number $n_q$ of 
generation-$(q+1)$ avalanches within a given generation-$q$ avalanche. 

Our basic observation is that we can derive a recursion relation, relating the distribution 
functions ${\cal P}_q(N)$ and ${\cal P}_{q+1}(N)$ of the total passing charge ($N^{(q)}$ and $N^{(q+1)}$, 
respectively) per avalanche for subsequent generations. This recursion follows from the obvious facts that 
$N^{(q)}=\sum_{j=1}^{n_q} N^{(q+1)}_j$ and that different avalanches of a given generation are 
statistically independent.  
By analogy with the derivation of Eq.\ (\ref{general}), we thus obtain
\begin{equation}
\label{recursive}
{\cal P}_q(N) =  \int \frac{d\alpha}{2\pi} e^{i\alpha N} \sum_{n} 
\left[\tilde {\cal P}_{q+1}(\alpha)\right]^{n} p_q(n),
\end{equation}
where $p_q(n)$ denotes the distribution function of $n_q$.
To proceed further one has to specify the form of the distribution $p_q(n)$.
This distribution is governed by the {\it microscopic} characteristics of the 
Franck-Condon transitions. We demonstrate below that 
$p_q(n)=(1/{\overline n}_q) \exp(-n/{\overline n}_q)$.
Upon substituting this form into Eq.\ (\ref{recursive}), the summation
over $n$ on the RHS can be easily performed and we obtain, after a Fourier transform 
of both sides,
\begin{equation}
   \tilde {\cal P}_q(\alpha) = \frac{1}{\overline n_q}\frac{1}{1- \tilde {\cal P}_{q+1}(\alpha) \exp(-1/\overline n_q ) }.
\label{con}
\end{equation}
The distribution ${\cal P}_q$ can 
now be obtained from this equation by writing its general solution as 
$\tilde {\cal P}_q(\alpha) = [1 + i\alpha \overline N^{(q)} + c_q (\alpha \overline 
N^{(q)})^2 +\ldots]^{-1}$. Inserting this into Eq.\ (\ref{con}), we find that
the numerical coefficients $c_q$ flow to zero with $q$ by virtue of the small parameter $1/\overline  
n_q$. Thus, the solution $\tilde {\cal P}_q(\alpha) = [1 + i\alpha \overline N^{(q)}]^{-1}$
with Fourier transform ${\cal P}_q(N)=\theta(N) \exp(-N/\overline N^{(q)})/\overline N^{(q)}$
can be viewed as a fixed point of the recursion equation Eq.\ (\ref{recursive}) and since 
$\overline N^{(q)} = \overline n_q\overline N^{(q+1)}$, self-similarity is obeyed asymptotically. 
The existence of this fixed-point solution can be viewed as a consequence of 
remark ({\em i}) which implies that up to rescalings, the distribution functions ${\cal P}_q(N)$ have 
the same functional form for {\em all} $q$. Fig.\ \ref{montecarlo}
numerically confirms this result for three different generations. 

\begin{figure}[t]
\includegraphics[viewport=50 600 250 745,width=0.8\columnwidth,clip]{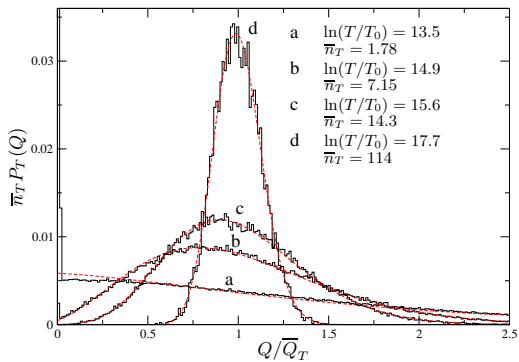}
\caption{(Color online) Evolution of full counting statistics $P_T(Q)$ for four different time intervals $T$ with $\lambda=4.0$ and $eV=3\hbar\omega_v$.
The MC data (solid lines) are in excellent agreement with the analytical full counting statistics, Eq.\ (\ref{main}), (dashed lines), with $\overline{N}^{(0)}=91.2$ and $\overline{n}_T=2.4\cdot10^{-6}T/T_0$ (no fit), as well as $T_0=(2\pi\rho t_0^2/\hbar)^{-1}$.}
\label{fullcount}
\end{figure}

With ${\cal P}_q(N)$ established, we obtain the counting
statistics by substituting $\tilde{\cal P}_0(\alpha)=
(1+i\alpha\overline N^{(0)})^{-1}$
into Eq.\ (\ref{similar}) and performing the integral.
This yields 
\begin{equation}
\label{main}
  P_T(Q) = e^{-{\overline n}_T} \delta(Q) +  
  e^{-\frac{Q}{ {\overline N^{(0)}}}-{\overline n}_T} 
  \sqrt{ 
  \frac{{\overline n}_T } {{\overline N}^{(0)} Q}
  }\, 
   I_1\!\left(\sqrt{\frac{4{\overline n}_T Q}{{\overline N}^{(0)}}} \right).
\end{equation}
Here $I_1(z)$ denotes a modified Bessel function. Eq.\ (\ref{main}) is our central
result. It is nicely confirmed by our MC results, as 
shown in Fig.\ \ref{fullcount}, and describes the evolution of the counting statistics between the following two 
transparent limits.
({\em i}) Short times, ${\overline n}_T=T/\langle t_i \rangle \ll 1$: Using the expansion 
$I_1(z)\approx z/2$ for $z\ll 1$ we obtain from Eq.\ (\ref{main})
\begin{equation}
\label{short}
P_T(Q)\simeq e^{-{\overline n}_T}\left[\delta(Q)+({\overline n}_T/\overline N^{(0)})
         e^{-{Q/ {\overline N}^{(0)}}}\right],
\end{equation}
Typically only a few electrons are transmitted through a molecule. The long tail described by the second term 
in Eq.\ (\ref{short}) arises from realizations where an avalanche occurs within the time $T$ and reflects the 
spread of charge within a single avalanche. 
({\em ii}) Long times, ${\overline n}_T \gg 1$: Upon substituting the large-$z$ asymptote of $I_1(z)$ 
into Eq.\ (\ref{main}), it is easy to see that the second term has a sharp maximum centered at 
$Q={\overline n}_T\overline N^{(0)}$, which is the average charge passed through the molecule after a large 
number of avalanches. Expansion of the exponent around the maximum yields the Gaussian
\begin{equation}
\label{long}
P_T(Q) \simeq (\sqrt{2\pi}\sigma_Q)^{-1}\exp\left[-{\left(Q-{\overline n}_T
\overline N^{(0)}\right)^2}/{2\sigma_Q^2}\right] 
\end{equation}
with a width $\sigma_Q=\left(2{\overline n}_T\right)^{1/2}\overline N^{(0)}$. This width is {\em twice} the width 
expected from the fluctuations of the waiting times. This enhanced broadening is due to fluctuations 
of the charge passed per avalanche. These additional fluctuations also manifest themselves in the noise 
characteristics of transport as analyzed below. 

{\it Microscopic derivation of $p_q(n)$.}---The distribution $p_q(n)$
is obtained by averaging the Poisson distribution of $n$ for a {\em given} 
avalanche duration over the distribution of {\em durations}.
On  microscopic grounds, the latter distribution is a simple 
exponent, which immediately transforms into $p_q(n)=(1/{\overline n}_q) \exp(-n/{\overline n}_q)$ 
since $\overline n_q$ is large.

To see this, we note that the duration $\tau^{(q)}$ of a generation-$q$ avalanche is determined by the 
waiting times in the vibrational state $q+1$ since the durations of intermittent higher-generation 
avalanches can be neglected. Two processes terminate a generation-$q$ avalanche: a direct 
transition from $q+1$ to $q$ or a transition back to $q$ during a generation-$(q+1)$ avalanche. Denoting the 
total rate for both processes by $\Gamma_q$, we obtain an exponential distribution of durations
$\Gamma_q \exp(-\Gamma_q\tau^{(q)})$. 

{\it Noise spectrum $S(\omega)$ of avalanche-type transport.}---We first derive a general expression for 
$S(\omega)$ assuming arbitrary distributions ${\cal P}_0(N)$ of the avalanche magnitudes and $W(t)$ of the 
waiting times. For frequencies smaller than $1/\tau^{(0)}$, we have $f_i^{(0)}(t)\simeq N_i\delta(t)$ in Eq.\ 
(\ref{current}). Using Fourier representations of the $\delta$ functions 
and averaging over the $t_i$ and $N_i$, the average current becomes
\begin{equation}
    \langle I(T)\rangle = \langle N_i\rangle \int\frac{d\alpha}{2\pi}  e^{i\alpha T} \frac{\tilde W(\alpha)}{
    1 - \tilde W(\alpha)} ,
\label{avcur}
\end{equation}
where $\tilde W(\alpha)=\langle \exp(-i\alpha t_i)\rangle_{t_i}$ denotes the Fourier transform of $W(t)$.
In the long-time limit, only small values of $\alpha$ contribute to the integral in Eq.\ (\ref{avcur}) so that
we can use the expansion $\tilde W(\alpha) = 1 -i\alpha \langle t_i\rangle - (1/2) \alpha^2
\langle t_i^2 \rangle + \ldots$.  Inserting this expansion into Eq.\ (\ref{avcur}), keeping only the leading 
order in $\alpha$ and performing the contour integration over $\alpha$, we recover the obvious result 
$\langle I(T)\rangle = \langle N_i\rangle /\langle t_i\rangle$. 

Similarly, we express the current-current correlator as
\begin{widetext}
\begin{eqnarray}
   &&\langle I(T_1) I(T_2)\rangle = \int \frac{d\alpha}{2\pi}\frac{d\beta}{2\pi} e^{-i\alpha T_1 - i\beta T_2}
      \left\langle [N_1 e^{i\alpha t_1}+N_2 e^{i\alpha (t_1+t_2)}+\ldots ][N_1 e^{i\beta t_1}+N_2 e^{i\beta 
      (t_1+t_2)}+\ldots ]\right\rangle \nonumber\\
       &&\,\,\,\,= \int \frac{d\alpha}{ 2\pi}\frac{d\beta}{ 2\pi} e^{-i\alpha T_1 -i\beta T_2}
         \left\{ \frac{\langle N_i\rangle^2 \tilde W(\alpha+\beta)}{
      1 - \tilde W(\alpha+\beta)}\left( \frac{1}{
    1 - \tilde W(\alpha)} + \frac{1}{
    1 - \tilde W(\beta)} -1 \right)\right.
      + \left. \frac{(\langle N_i^2\rangle -\langle N_i\rangle^2) \tilde W(\alpha+\beta)}{
    1 - \tilde W(\alpha+\beta)}\right\}.
\label{corcur}
\end{eqnarray}
\end{widetext}
The last equality follows upon term-by-term averaging and 
resummation of the series. To access the limit of long times $T=(T_1+T_2)/2$, we introduce
$\omega = (\alpha -\beta)/2$ and $\Omega = \alpha +\beta$. Then, the exponent in the integrand 
in Eq.\ (\ref{corcur}) assumes the form $\exp(i\omega\tau - i\Omega T)$ with  
$\tau=T_2-T_1$. The limit $T\to\infty$ can now be taken in analogy with the 
derivation of $\langle I(T)\rangle$ above. The integrand can be directly
identified with the noise 
spectrum $S(\omega)$, so that
\begin{align}\nonumber
   S(\omega)&= \frac{2}{ \langle t_i\rangle} \bigg\{ \langle N_i\rangle^2 \left[ \frac{1}{
    1 - \tilde W(\omega)} + \frac{1}{
    1 - \tilde W(-\omega)} -1 \right]\\
    &\qquad +(\langle N_i^2\rangle -\langle N_i\rangle^2) \bigg\}.
\label{noisespectrum}
\end{align}
Taking the zero-frequency limit requires one to keep terms of order $\omega^2$ in the 
expansion of $\tilde W(\pm \omega)$. In this way, the Fano factor $F=S(\omega=0)/2e\langle I\rangle$ 
becomes
\begin{equation}
  F=\langle N_i\rangle \frac{\langle t_i^2\rangle-\langle t_i\rangle^2}{\langle t_i\rangle^2}
          +\frac{\langle N_i^2\rangle-\langle N_i\rangle^2}{ \langle N_i\rangle}.
\label{fanogen}
\end{equation}
This equation allows for a transparent interpretation: Noise originates 
from two sources, namely the fluctuations in the intervals between avalanches and the fluctuations in the 
transmitted charge per avalanche. In the conventional situation where $N_i=1$ for all $i$, the Fano factor 
is given by the fluctuations of the waiting times $t_i$ for a transition in which an electron passes either 
directly or sequentially from the left to the right lead. For example, for transport through a symmetric 
junction in the Coulomb-blockade regime, one immediately recovers $F=5/9$ \cite{Nazarov} when taking into 
account that the rates of entering and leaving the dot are related as 2:1 due to spin. 
For the specific distributions adopted in our model, both terms in Eq.\ (\ref{fanogen}) contribute equally, 
and the Fano factor reduces to $F=2{\overline N}^{(0)}$, which, in agreement with Eq.\ (\ref{long}), is twice the 
value expected for a fixed magnitude of avalanches. This is confirmed by numerical results.

For frequencies larger than $1/\tau^{(0)}$, the  ``fine structure" 
of the avalanches described by the functions $f_i^{(0)}$ in Eq.\ (\ref{current})
must be taken into account. This fine structure can be incorporated into
the noise spectrum Eq.\ (\ref{noisespectrum}) by replacing $\langle N_i\rangle^2$ by $\langle \tilde f(\alpha)\rangle\langle \tilde f(\beta)\rangle$
and $\langle N_i^2\rangle -\langle N_i\rangle^2$ by $\langle \tilde f(\alpha)
\tilde f(\beta)\rangle - \langle \tilde f(\alpha)\rangle\langle \tilde f(\beta)\rangle$, where $\tilde f(\alpha)$
denotes the Fourier transform.
Explicitly employing the Poisson distribution of the waiting times leads to 
the remarkable simplification
$[1-\tilde W(\omega)]^{-1}+[1-\tilde W(-\omega)]^{-1} -1 = 1$. In this way, we obtain
\begin{equation}
   S(\omega) = \frac{2}{ \langle t_i\rangle} \langle \tilde f(\omega)\tilde f(-\omega)\rangle.
\label{Somega}
\end{equation}
For frequencies of order $\omega\simeq 1/\tau_0$ (where $\tau_q$ denotes the average waiting time 
$\langle t_i\rangle$ at level $q$ of the hierarchy), we can ignore the fine structure of the avalanche
and replace $\tilde f(\omega) = N_i^{(0)}$. Thus, we find $S(\omega) = 2 \langle [N_i^{(0)}]^2\rangle /
\tau_0$. At higher frequencies $\omega \simeq 1/\tau_1$, the function $f(t)$ is resolved into avalanches
of generation $q=1$. Then, we can write $\langle\tilde f(\omega)\tilde f (-\omega)\rangle
= \int dT\int d\tau e^{i\omega\tau} \langle f(T+\tau/2)f(T-\tau/2)\rangle$. Up to the integral over
$T$, this expression is analogous to $S(\omega)$ itself, with zeroth-generation quantities replaced by
corresponding $q=1$ quantities. For the frequencies of interest, we therefore find $S(\omega) = (2/\tau_0)(\tau^{(0)}\langle
[N_i^{(1)}]^2\rangle/\tau_1)$. Using the obvious relations $\tau^{(0)}=\tau_1 {\overline n}_0$ and 
$\overline N^{(0)}=\overline n_0 \overline N^{(1)}$ and generalizing to arbitrary $q$, we find
\begin{equation}
   S_{q+1} = \frac{{\overline N}^{(q+1)}}{ {\overline N}^{(q)}} S_q. 
   \label{Shierarchy}
\end{equation}
Here, we define $S_q = S(\omega \simeq 1/\tau_q)$ so that Eq.\ (\ref{Shierarchy}) provides a rule for 
extending the noise spectrum to progressively higher frequencies.

The essential {\it microscopic} inputs are the ratios $\tau_{q+1}/\tau_q$ and ${\overline N}^{(q+1)}/
{\overline N}^{(q)}$. Both ratios are determined by overlaps of displaced vibrational wavefunctions
\cite{Koch}. 
The rate $1/\tau_q$ is dominated by the transition $q\to q+1$. Thus, it involves the overlap of 
{\it neighboring} harmonic oscillator states. By contrast, ${\overline N}^{(q)}$ is inversely proportional 
to the transition rate from a highly excited phonon levels to the $q$th vibrational level.
The difference between these two rates is thus that the first involves four wavefunctions with 
index of order $q$, while the second involves only two. As a result, we can immediately establish 
from a quasiclassical evaluation of the matrix elements that $\tau_q/({\overline N}^{(q)})^2$ is essentially
independent of $q$. With this input, we conclude that $S(\omega) \sim \omega^{-\alpha}$
with exponent $\alpha=1/2$. Since the noise power does not depend sensitively on $\omega$ in finite 
intervals around $ 1/\tau_q$, this power law should be superimposed with
steplike features in $S(\omega)$. These conclusions agree 
with numerical simulations (see Ref.\ \cite{Koch}) over several orders of magnitude in frequency.

{\it Conclusions.}---Our complete analytical description for the full-counting statistics and the 
frequency-dependent noise power of self-similar avalanche-type transport was made possible 
by the fact that current flow is essentially unidirectional. We emphasize that our arguments are
quite general, with rather limited microscopic input, making our results potentially applicable 
far beyond the particular realization \cite{Koch} of avalanche-type transport considered in the 
numerical simulations. Finally, we remark that direct vibrational relaxation (with rate $\gamma_\text{rel}$), 
neglected so far, only gradually suppresses avalanche-type transport. Indeed, 
one readily argues that $\overline N^{(0)}\sim 1/\gamma_\text{rel}$ over a wide range of relaxation times, 
leading to $S(0)\sim 1/\gamma_\text{rel}^2$.

This work was supported by NSF-DAAD Grant INT-0231010, Sfb 658, and Studienstiftung des deutschen 
Volkes.


\begin{thebibliography} {50}
\bibitem{terminology} Strong electron-phonon coupling means $\lambda\gg 1$,
where $\lambda$ is the shift of the equilibrium 
nuclear normal mode between neutral molecule and ion, in units of the 
spatial extent of the vibrational ground-state wavefunction.
\bibitem{cutoff} The hierarchy breaks off once the transferred charge per avalanche is of order 1.
\end{thebibliography}
\end{document}